\documentclass[usenatbib]{basi}
\usepackage[T1]{fontenc}
\usepackage[british]{babel}
\usepackage[varg]{txfonts}
\usepackage{rotating}
\usepackage{dcolumn}
\usepackage{enumitem}
\setlist[itemize]{noitemsep}
\setcitestyle{numbers}
\begin{document}
\title[]{Probing obscured, high redshift galaxies using deep P-band continuum imaging with GMRT}
\author[Y.~Wadadekar et~al.]%
       {Yogesh Wadadekar$^1$\thanks{email: \texttt{yogesh@ncra.tifr.res.in}},
       Sandeep Sirothia$^{1}$,
       C. H. Ishwara-Chandra$^{1}$,
       \newauthor
       Veeresh Singh$^{2}$, 
       Alexandre Beelen$^{2}$ and Alain Omont$^3$ \\ 
       $^1$National Centre for Radio Astrophysics, TIFR, Post Bag 3, Ganeshkhind, Pune 411007, India\\
       $^2$ Institut d'Astrophysique Spatiale, B\^{a}t. 121, Universit\'{e} Paris-Sud, 91405 Orsay Cedex, France\\ 
       $^3$UPMC Univ Paris 06 and CNRS, UMR 7095, Institut d'Astrophysique de Paris, F-75014, Paris, France}
\pubyear{2014}
\volume{00}
\pagerange{\pageref{firstpage}--\pageref{lastpage}}

\date{Received --- ; accepted ---}

\maketitle
\label{firstpage}

\begin{abstract}

We have carried out a deep (150 micro Jy rms) P-band, continuum
imaging survey of about 40 square degrees of sky in the XMM-LSS,
Lockman Hole and ELAIS-N1 fields with the GMRT. Our deep radio data,
combined with deep archival observations in the X-ray (XMM/Chandra),
optical (SDSS, CFHTLS), near-infrared (UKIDSS, VISTA/VIDEO),
mid-infrared (Spitzer/SWIRE, Spitzer/SERVS) and far-infrared
(Spitzer/SWIRE, Herschel/HerMES) will enable us to obtain an accurate
census of star-forming and active galaxies out to $z\sim 2$.  This
panchromatic coverage enables accurate determination of photometric
redshifts and accurate modeling of the spectral energy
distribution. We are using our large, merged photometric catalog of
over 10000 galaxies to pursue a number of science goals.

\end{abstract}


\section{Introduction}\label{s:intro}

Understanding the mass assembly history of galaxies remains a major
challenge in astrophysics.  This is because a major fraction of galaxy
assembly happens at high redshifts, in intense bursts of star
formation as well as black hole accretion. Therefore, disentangling
starburst and AGN activity, and understanding why the peak in the
comoving luminosity density of star formation coincides with that from
AGN activity at $z\sim2$, is vital for measuring the stellar/BH mass
buildup. Since most of the energy from these activities is absorbed by
dust and then re-radiated in the rest-frame infrared, comprehensive
multi-wavelength studies are required to achieve this, with emphasis
on far-IR/submillimeter observations which probe the peak in the IR
emission from moderately warm dust, supplemented by radio surveys
which can probe the inner regions of star formation and AGN activity,
without being adversely affected by dust obscuration.

The last couple of years have seen the arrival of revolutionary
FIR/submm surveys, on a large new FIR telescope -- {\em
  Herschel}. Among those, the Herschel Multi-tiered Extragalactic
Survey (HerMES\footnote{http://www.hermes.sussex.ac.uk}) is the
largest (900\,hr) guaranteed time program. It aims to chart the
evolution of galaxies through cosmic history via a set of nested (wide
and shallow as well as deep pencil-beam) surveys over twelve well
studied areas of sky, in particular, those extensively studied by {\it
  Spitzer} at 24 and 70 $\mu$m, such as the ELAIS-N1, Spitzer FLS,
CDFS, GOODS-N, Lockman and XMM-LSS fields (total area coverage of
$\sim$110 deg$^2$ plus 270 deg$^2$ of shallow survey).  

As demonstrated by the results already published in about 40 refereed
papers (see the HerMES website for a list of these papers), HerMES has
provided new insight on distant dusty galaxies and AGNs. It seems that
previous phenomenological galaxy populations need revision and it is
now anticipated that HerMES will be able to catalogue over 100,000
galaxies with $>5\sigma$ detections at 250 $\mu$m. HerMES will
constitute a lasting legacy to the community, providing an essential
complement to multiwavelength surveys in the same fields and providing
targets for follow-up using many facilities for many years to come.

\section{Our radio followup}

These dramatically improved FIR data have only one limitation -- their
relatively poor resolution (18,25,36 arcsec FWHM for Herschel/SPIRE at
250, 350, 500 $\mu$m resp.). Radio interferometric observations with
the GMRT at 325 MHz provide accurate (1.5 arcsec) astrometry, even for
faint sources, allowing matches to data at other wavelengths, and
thereby redshift determination. With deep radio data, radio
counterparts can be identified for many FIR sources because of the
strong and tight radio-FIR correlation; FIR bright star-forming
galaxies are invariably bright in the radio. We use the 325 MHz band
of GMRT for our observations because it provides an optimal
combination of good sensitivity (even more so for steep spectrum
sources), moderate RFI, large field of view (half power beamwidth of
83 arcmin) and adequate resolution (beam size of $\sim$10 arcsec and
positional uncertainty of $\sim1.5$ arcsec even for the fainter
sources).

\begin{figure}
\centerline{\includegraphics[width=11cm]{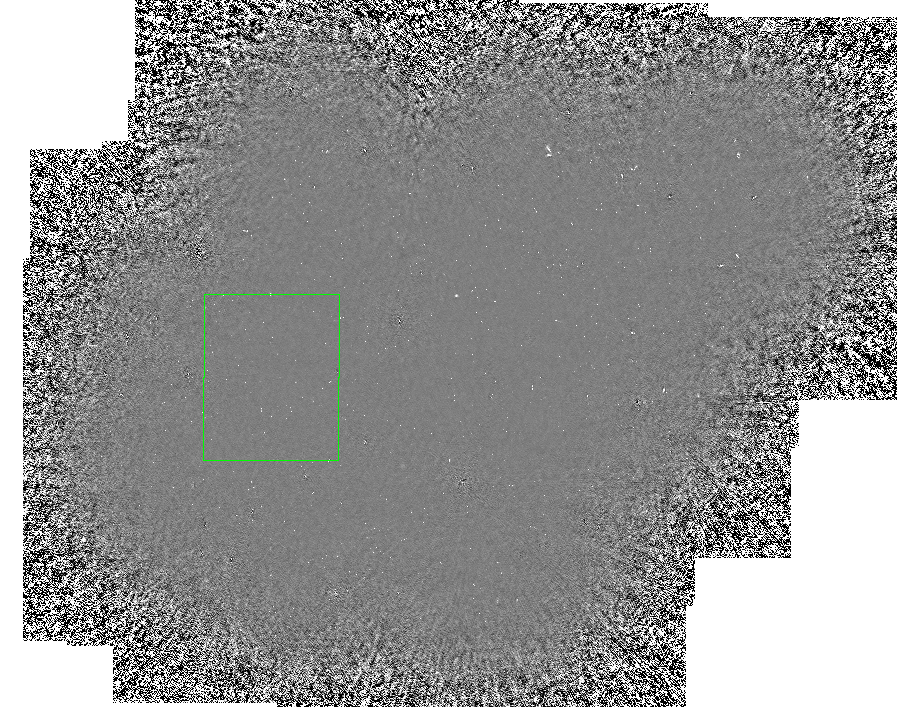}}
\caption{Our 325 Mhz radio map of the XMMLSS field.  The larger blue rectangle (only partially shown) is the area covered by the CFHTLS-W1 field. The small blue square corresponds to the CFHTLS-D1 field which also has deep radio observations with VLA/GMRT. Overplotted as the  green rectangle is the coverage of the VISTA Video XMM-3 field. The jagged yellow rectangle is the coverage of the Spitzer/SERVS survey.  The inclined yellow square is the area with multiband observations from the Spitzer-SWIRE survey.The orange contour shows the coverage of the Herschel/HerMES observations while the red ellipse is the coverage of the deep 1.4 GHz VLA observations by Simpson et al (2006). Over $10^5$ galaxy and quasar spectra are also available over our map.\label{f:xmmlsscoverage}}
\end{figure}

The GMRT field of view at 325 MHz is optimal for followup of the
wide-shallow component of HerMES. We have observed three such fields -
XMMLSS, Lockman Hole and ELAIS-N1 (40 deg$^2$ in total) to a depth of
$\sim150 \mu$Jy rms. In each field, we have covered the full area covered
by the Herschel/HerMES and Spitzer/SWIRE surveys to maximise the
scientific returns and legacy value. These well studied fields have a
plethora of archival data (both imaging and spectroscopy) available in
most bands from X-ray through radio (see Fig.~\ref{f:xmmlsscoverage})

\section{Ongoing investigations}

With this multiwavelength dataset, a number of scientific
investigations can be carried out. We are currently working on the
following projects.

\begin{itemize}

\item identify candidate high-$z$ powerful radio galaxies using the
  $K-z$ relation and radio spectral index measurements and study their
  properties (see Singh et al. this volume).

\item use image stacking of the HerMES and GMRT data explore the
  radio-FIR correllation in normal galaxies with $0 < z < 1.2$ well
  below the detection limit of the HerMES or GMRT images.

\item separate starburst and AGN (Ibar et al. 2010) and identify
  powerful obscured AGN using e.g., the methods of Seymour et
  al. (2008, 2009)

\item explore the radio/far-IR correlation in starbursts at high
  redshift. A recent study from the Herschel-ATLAS survey indicates
  that the correlation does not change over the redshift range $0 < z
  < 0.5$ (Jarvis et al. 2010). 

\item combine GMRT, HerMES and  SERVS data (which probes rest-frame
  1-3$\mu$m) to comprehensively test models of radio-mode feedback as a
  function of epoch and galaxy mass, with the GMRT data measuring
  feedback into the IGM by radio jets, HerMES data the current star
  formation rate and Spitzer data the stellar mass.

\end{itemize}

\begin{figure}
\centerline{\includegraphics[width=11cm]{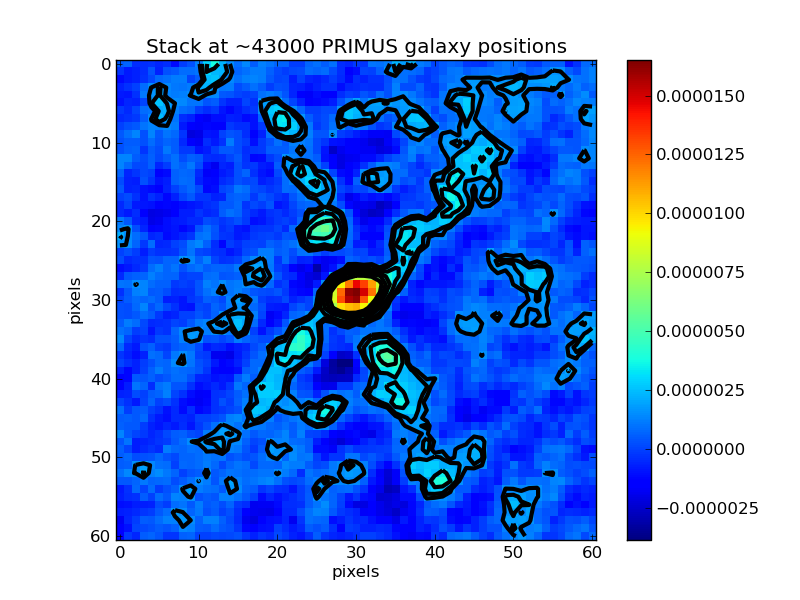}}
\caption{Median 325 MHz stack in the radio of $\sim 43000$ PRIMUS survey galaxies in the XMM-LSS field. The peak flux is only $~15 \mu$Jy/beam and the image rms is $\sim1 \mu$Jy/beam. The X shaped structure is the dirty beam of the GMRT from sub CLEAN-threshold sources contributing to the image stack.\label{f:hermesstack}}

\end{figure}

\section*{Acknowledgements}

We gratefully acknowledge support from the Indo-French Centre for Promotion of
Advanced Scientific Research (CEFIPRA) under Project 4404-3 titled ``Distant obscured galaxies from GMRT and Herschel''. 


\end{document}